\documentclass[pre,3p,twocolumn,floatfix]{revtex4-1}
\usepackage{graphicx,epsfig,array,pstricks,pstricks-add}
\usepackage{color,graphicx,amssymb,amsmath,hyperref,natbib}
\begin{document}
% Computer Physics Communications
%JOURNAL OF COMPUTATIONAL PHYSICS
\title{Parallelized Traveling Cluster Approximation to Study Numerically\\ 
Spin-Fermion Models on Large Lattices}
%\author[utk]{Anamitra Mukherjee}
%\ead{amukher5@utk.edu}
%\author[utk]{Niravkumar D. Patel}
%\author[utk]{Chris Bishop}
%\author[utk,ornl]{Elbio Dagotto}
%\address[utk]{Department of Physics and Astronomy, The University of Tennessee, Knoxville, Tennessee 37996, USA} 
%\address[ornl]{Materials Science and Technology Division, Oak Ridge National Laboratory, Oak Ridge, Tennessee 37831, USA}

\author{Anamitra Mukherjee$^1$}
\author{Niravkumar D. Patel$^1$}
\author{Chris Bishop$^1$}
\author{Elbio Dagotto$^{1,2}$}
\affiliation{$^1$Department of Physics and Astronomy, The University of Tennessee, Knoxville, Tennessee 37996, USA}
\affiliation{$^2$Materials Science and Technology Division, Oak Ridge National Laboratory, Oak Ridge, Tennessee 37831, USA}

\date{\today} 
\begin{abstract}

Lattice spin-fermion models are important to study correlated 
systems where quantum dynamics allows for a separation between 
slow and fast degrees of freedom. The fast degrees of freedom 
are treated quantum mechanically while the slow variables, 
generically refereed to as the ``spins'', are treated classically. 
At present, exact diagonalization coupled with classical Monte Carlo (ED+MC) 
is extensively used to solve numerically a general class of lattice spin-fermion 
problems. In this common setup, the classical variables (spins) 
are treated via the standard MC method while the fermion problem is 
solved by exact diagonalization. The ``Traveling Cluster 
Approximation'' (TCA) is a real space variant of the ED+MC method 
that allows to solve spin-fermion problems on lattice sizes with up to 
$10^3$ sites. In this publication, we present a novel reorganization 
of the TCA algorithm in a  manner that can be 
efficiently parallelized. This allows us to solve generic spin-fermion 
models easily on $10^4$ lattice sites and with some effort on $10^5$ lattice 
sites, representing the record lattice sizes studied for this family of models. 
\end{abstract}
\maketitle

%------------------- I   INTRODUCTION -----------
\section{Introduction}

The rich physical properties displayed 
by many materials arise from strong correlations among 
multiple degrees of freedom \cite{dagotto-science, nagaosa_science}. 
Studying theoretically these materials has been a long standing 
challenge for materials theory since treating those coupled 
multiple degrees of freedom (DOF), such as the 
spin, charge, orbital, and lattice, on equal footing in a model Hamiltonian 
calculation is extremely difficult. 
As a consequence, accurate approximations 
that render such complex problems more tractable have always been of 
considerable interest. Dynamical Mean Field Theory \cite{dmft-review}, 
Determinant Quantum Monte Carlo \cite{dqmc,WhitePRB1989,paiva10}, and the Density 
Matrix Renormalization Group \cite{dmrg-review} are some of those 
approximations that have led to important insights into 
the physics of correlated materials. Another useful 
approximation is to exploit the relative slow dynamics of some 
degrees of freedom as compared to others. As discussed below, 
this approach allows for the modeling of some complex materials 
with relative ease and on reasonably larger lattice sizes.

In materials such as the manganites \cite{mang-book,mang-rev-tokura}, 
double perovskites \cite{dp-rev-serrate}, rare earth 
nickelates \cite{nick-1,nick-2}, and others, the slow and fast 
separation is a good approximation. For example, in the manganites, 
the electrons in the $e_g$ orbitals have faster dynamics as compared 
to the dynamics of the localized $t_{2g}$ electrons and also compared to
the Jahn-Teller and breathing mode phonons \cite{mang-book,hotta-review}. 
This allows for a separation between ``fast'' and ``slow'' DOF. 
The quantum+classical approach treats the slow variables in the 
strict adiabatic limit, \textit{i.e.}, classically. Generically the 
slow variables that are considered classically are called ``spins'' 
and for this reason the models are commonly 
referred to as ``spin-fermion'' models.

The main advantage of this approximation is that the original 
fully interacting quantum many body problem can be mapped into a 
problem of noninteracting fermions coupled with, in general, spatially 
fluctuating classical fields. In the past, such classical+quantum 
approaches have been extensively used. Some well 
known methods in this context include the study of 
electron-phonon systems \cite{bo-1, bo-2}, 
the Born-Oppenheimer approximation \cite{bo-3}, 
and the Car-Parrinello method \cite{cp-1}. Spin-fermion models 
for the manganites \cite{hotta-review,yunoki,science99,tca-2,dong1,dong2,liang11,sen}, double perovskites \cite{dp-th-1, dp-th-2}, nickelates \cite{nick-th-1, nick-th-2}, copper based high temperature
superconductors \cite{buhler1,buhler2,mora1,mora2,mora3}, BCS superconductors \cite{BdG0,BdG1,BdG2,BdG3}, and the recently 
discovered iron superconductors \cite{pn-ku-1,pn-spin-fermion, pnictide-rev,shuhua-pn-1,shuhua-pn-2,shuhua-pn-3} 
have all exploited the slow and fast variables to considerable success. 

Solving such spin-fermion models entails the search for the
optimal configurations of the classical DOF that minimize 
the free energy. To achieve this goal, first the fermionic 
problem is diagonalized for a fixed configuration of the 
classical DOF and the energy is computed. The classical 
variables are then updated and the energy is recalculated 
in the updated background. The updates are accepted or rejected 
via the Metropolis algorithm. Finally the procedure is repeated 
until thermal equilibrium is reached and observables can be
measured with reasonable accuracy. 

This Exact Diagonalization + Monte Carlo (ED+MC) approach is free 
from the ``sign problems'' suffered by Quantum Monte Carlo 
methods and it can also include the study of 
long range spatial correlations unlike simple DMFT approaches. 
Over the years the ED+MC method has enjoyed considerable 
success in understanding correlated materials phenomena 
where the separation of  slow and fast DOF is possible \cite{hotta-review}. 
However, even after the considerable numerical simplification due to the 
quantum+classical treatment, the ED for the fermion problem still 
has to be carried out at every update of the classical fields resulting 
in thousands of diagonalizations at every temperature where
the calculation is performed. Furthermore, the simulated annealing 
from high to low temperatures, which is often required to avoid being trapped
in metastable states, requires sequential temperature steps. 
All these steps amount to a prohibitively large number of diagonalizations to be 
performed in a standard ED+MC calculation. This typically 
limits the accessible lattice or system sizes that can be 
solved using ED+MC to $\sim 10^2$.

The ability to solve such spin-fermion problems on larger systems 
is needed to address issues such as large length scale 
phase separation tendencies, to achieve accurate estimations 
of thermodynamic order and transport properties with small 
size effects, and to be able to perform reliable 
finite-size scaling analysis. Moreover, studies 
of the iron based superconductors have pointed out the 
need to study spin-fermion and Hubbard-like models incorporating
multiple orbitals \cite{pn-raghu,johnston,stewart,natphys12,RMP13}. This task is challenging 
even on small system sizes due to large Hilbert spaces. In these 
regards the simple ``Traveling cluster approximation'' (TCA)~\cite{kumar} 
is an important step forward as it allows access to system 
of $\sim 10^3$ sites. This approximation is discussed below. 
In this publication, we present an alternate way to organize 
the TCA algorithm that allows for massive parallelization 
of the method. As a result, the calculation of spin-fermion 
models can now be performed on system sizes up to $\sim 10^5$ sites. Additionally, as discussed later, in the present generalization  very large traveling clusters can be used for the TCA calculation. The small size of the traveling clusters has, till now, remained a limitation of the TCA approach.

Below we describe the parallelization scheme and 
benchmarks that we developed. As discussed 
in the text, techniques of the nature developed here will be instrumental 
in addressing problems in multiband Hubbard models as well.

The paper is organized as follows. In section II, we explain 
the TCA technique and compare it with ED+MC. In section III, we 
discuss our approach for parallelizing the TCA algorithm. 
In section IV, we present benchmarking results and in section V we 
provide some physically relevant results for the one orbital 
Hubbard model both in two and three dimensions and compare 
them with existing literature. In sections VI and VII, 
we discuss some pertinent numerical issues and in section VIII, we present the conclusions of the manuscript.

\section{Traveling Cluster Approximation}

Let us begin by briefly discussing the basics of 
the ED+MC and TCA approaches. 

\textit{a. ED+MC :} As mentioned before, the spin-fermion 
model consists of a classical component and a quantum component. 
A Hamiltonian for spin-fermion models define the coupling 
between the classical DOF and the electrons and among 
the classical variables themselves. In usual ED+MC approaches, 
the classical variables at each site are updated one at a time 
and the energy of the system is calculated by diagonalizing 
the Hamiltonian and adding the classical contribution. This 
energy difference, before and after the update at a site, is 
used to accept or reject the proposed update. The process is 
then repeated over all of the sites visiting them either 
serially or randomly. This constitutes a single system ``sweep''. 
The combined algorithm of ED+MC is numerically rather costly, 
since the exact diagonalization must be performed at every step 
and the cost scales as $\mathcal{O}(N^3)$ with $N$ the number 
of lattice sites. Additionally, with a sequential system sweep, 
the cost of a Monte Carlo system sweep scales as $N^4$ at each 
temperature. 
\begin{figure}[t]
\centering{
\includegraphics[width=6cm, height=6cm, clip=true]{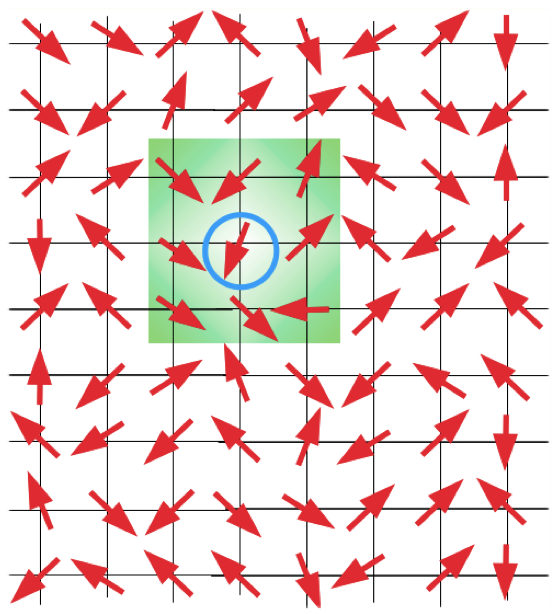}}
\caption{Two dimensional schematic of the TCA approach. Here a $N=8^2$  
lattice is displayed with classical DOF at each site, represented by 
the red arrows. The Hamiltonian defines the coupling of the classical 
spins with the itinerant electrons that are delocalized on 
the lattice. The TCA algorithm consists of proposing an update 
at a site (encircled in blue) for the classical spins. The update 
is accepted or rejected based on the energy of a cluster  
built around that site, indicated in green. Here the cluster 
size is $N_c=9$. In a system sweep, the above procedure is 
carried out by visiting each site of the system sequentially.}
\vspace{-0.0cm} 
\label{f-1}
\end{figure}

\textit{b. TCA scheme:} To reach reasonably large system sizes, a real space variant  \cite{kumar} of the ED+MC approach 
has been developed. As will be discussed below, this allows for
a linear scaling with the system size, $N$, as opposed to the 
$N^4$ scaling of the computational cost with ED+MC. 

\begin{figure}[t]
\centering{
\includegraphics[width=8cm, height=12cm, clip=true]{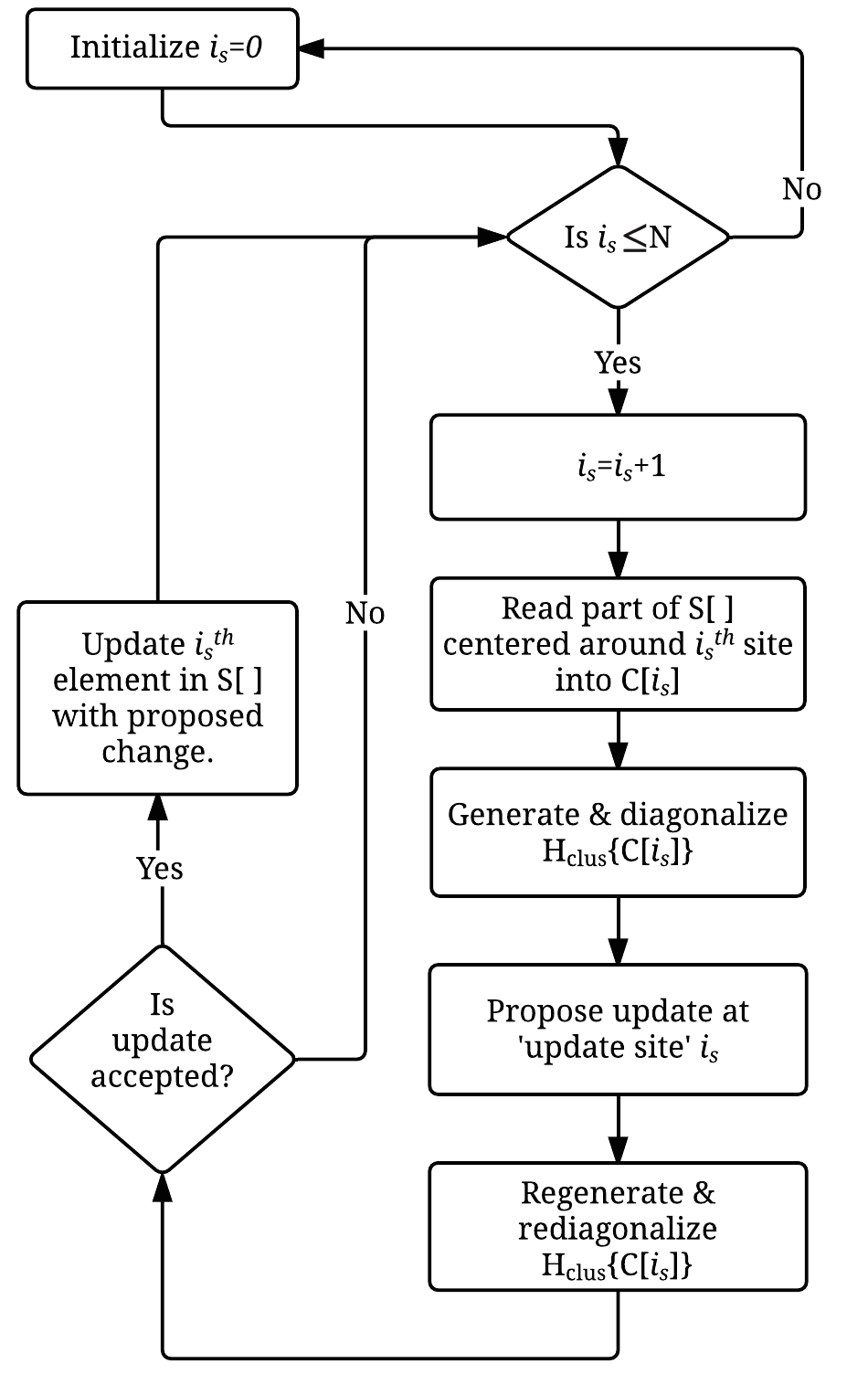}}
\caption{The flowchart for a single sequential Monte Carlo 
system sweep in the TCA approach. Here, we consider a lattice 
of $N$ sites. $S[~]$ is an array containing the classical DOF 
at all the $N$ sites. $C[i_s]$ is an array of size $N_c$ that 
reads in the classical DOF from $S[~]$ around the ``update site'' 
labeled by $i_s$. The loop over $i_s$ runs over all the $N$ sites 
of the system, ensuring that the cluster is built around all the 
sites sequentially and that an update is proposed at each site. 
The flowchart is discussed in the text.
}
\vspace{-0.0cm} 
\label{f-2}
\end{figure}

In the TCA scheme one defines a region (cluster) around the site 
where a MC update is attempted. The cluster has a linear dimension 
$L_c$. Then, in a two  dimensional square lattice, for example, 
the number of sites in the cluster is $N_c=L_c^2$. Such a cluster 
is shown in Fig.~\ref{f-1}. The cluster 
is built around a site called the ``update'' site that it is encircled 
in blue. In this example $N_c=9$. The key difference with ED+MC 
is that the proposed update is accepted or rejected on the basis 
of the energy difference of the cluster and \textit{not} the full 
system. As a result one needs to diagonalize only the cluster 
Hamiltonian which costs $N_c^3$ as opposed to $N^3$ for the 
full system diagonalization in ED+MC. 

The analytical basis for the approximation of using a smaller 
cluster for the annealing process lies in the principle of 
``nearsightedness'' of electronic matter, as discussed 
by W. Kohn \cite{locality-1, kohn-2}. Furthermore, the method 
has been extensively tested and benchmarked in numerical 
studies \cite{tca-1,tca-2}. Consequently, in this paper 
we will assume the validity of the approximation without 
further discussion. The above update scheme is sequentially employed
at every site of the system. The cluster of size 
$N_c$ is built around every site where the update 
is attempted, hence the name ``Traveling'' cluster approximation.
Thus, within TCA, and at each temperature, the computational 
cost of ED for a system with $N$ sites is $\mathcal{O}(N_{c}^{3})$ 
and the cost of a full sweep of the lattice is $NN^3_c$ or {\it linear} 
in $N$ as opposed to $N^{4}$. 
 
Many thousand MC system sweeps are performed at every temperature 
and for each temperature a large number of annealed classical 
configurations are stored. These are later used to construct 
and  diagonalize the full system if it is necessary for calculating 
the desired output quantities. They are also useful for studying 
correlations among the classical variables.

As discussed in the original TCA paper \cite{kumar}, the geometry 
of the cluster is chosen to be the same as the system. Furthermore, 
one has to impose periodic boundary conditions on the cluster while 
calculating energies. These conditions ensure that in the limit 
of cluster sizes  approaching the actual system size, the spectrum 
becomes identical. The periodically identified cluster can be 
considered to be an independent ensemble in contact with the 
full system where equilibrium is maintained in a grand canonical 
framework. An important aspect of this setup is that any site on 
the cluster can be chosen as the ``update site''. In Fig.~\ref{f-1} 
we show this update site to be equidistant from all the edges. 
However, any other site, for example, the one in the top left corner, 
is also a equivalently good choice. This equivalence has been tested 
in many numerical studies \cite{kumar,tca-1,tca-2} and we also 
checked numerically the same concept 
in the context of the Holstein model in section VII.

\textit{c. TCA flowchart:} We end this section with the TCA algorithm,  
and the corresponding flowchart is presented in Fig.~\ref{f-2}. In the 
flowchart the following nomenclature is used, and the same will be used 
for discussing the PTCA approach as well. We consider a system with $N$ 
sites. $S[~]$ is an array containing the classical variables at each site 
and it has the length $N$, assuming one classical variable per site. 
$H_{full}\{S[~]\}$ is the Hamiltonian for the full system, generated 
from the classical variables in $S[~]$. The array $C[i_s]$ is of length 
$N_c$ and it is the array holding the classical DOF at the cluster sites 
built around the $i_s^{th}$ site of the system. So this array reads the 
relevant part of $S[~]$. For example, in Fig.~\ref{f-1} $C[i_s]$ will 
read in, from $S[~]$, all classical variables that are at the lattice 
sites covered in the green square. From this setup the cluster Hamiltonian, 
$H_{clus}\{C[i_s]\}$, is constructed around the $i_s^{th}$ site. 
The $i_s^{th}$ site, where the update is proposed and around which 
the cluster is built, is referred to as the ``update site.'' 
With these notations, the following are the main steps explaining 
the TCA flowchart presented in Fig.~\ref{f-2}.

\begin{enumerate}
\item In a single Monte Carlo system sweep, the index $i_s$ loops 
over all $N$ sites of the full system. Around each of the sites, $i_s$, 
a cluster will be built one at a time, traveling sequentially, 
as $i_s$ sweeps over the full lattice.
\begin{enumerate}
\item $C[i_s]$ reads part of $S[~]$ around the site $i_s$.
\item $H_{clus}\{C[i_s]\}$ is generated and diagonalized.
\item The classical DOF is randomly modified at site $i_s$.
\item $H_{clus}\{S[~]\}$ is generated with the update and rediagonalized.
\item A Metropolis algorithm decides if the proposed update is accepted or not.
\item If accepted the $i_s^{th}$ element in $S[~]$ is changed to the updated value.
\end{enumerate}
\item The above process is repeated for all sites of the system.
\end{enumerate}

\section{Scheme for parallelization}

\begin{figure}
\centering{
\includegraphics[width=6.6cm, height=5cm, clip=true]{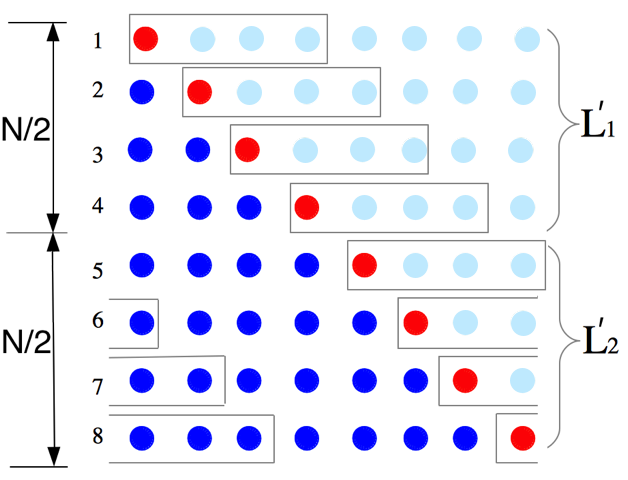}}
\caption{One dimensional example using $N=8$ and $N_c=4$. 
The light blue sites are the un-updated sites. The red site is 
chosen to be the site where the update is attempted (update site). 
The clusters are indicated by the boxes. The different rows present
 the cases of the cluster traveling sequentially from left to right 
during a single system sweep. The blue sites indicate the sites where 
an update has been attempted. The system and cluster both have 
periodic boundary conditions. For rows 1  through 5, 
the clusters are built based on the original (un-updated) 
classical variable configurations and can be constructed 
simultaneously instead of serially as in TCA. For rows 6 
through 8, the clusters require the results of the 
update attempts on sites 1 through 3, respectively. 
So these clusters have to wait till row 4 
and then can be built simultaneously.}
\vspace{-0.0cm} 
\label{f-3}
\end{figure}

\begin{figure*}[t]
\centering{
\includegraphics[width=17cm, height=16cm, clip=true]{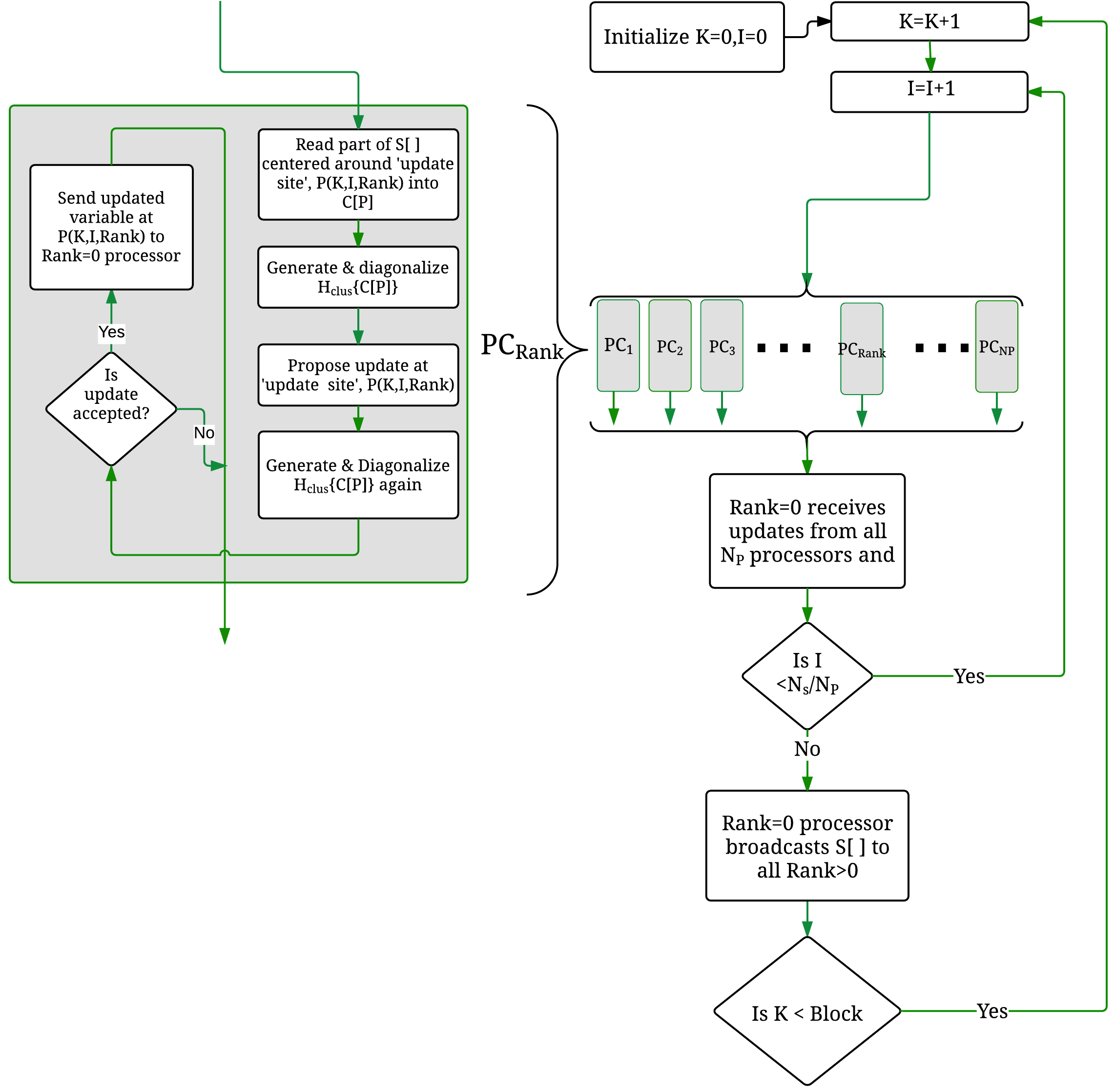}}
\caption{Parallelized TCA flowchart for a single system sweep. 
Unlike  Fig.~\ref{f-2}, the loop over the total number of sites of 
the system is split into two loops: the outer one runs over 
the number  of blocks, while the inner one runs from 1 to the ratio 
of the number of sites in a block ($N_S$) to the number 
of processors ($N_P$). The gray rectangles, labeled 
$PC_1$, $PC_2$, ..., $PC_{NP}$, are computed on processors 
with rank 1 through $N_P$, respectively. The steps in a 
typical gray block are displayed on the left. See text for discussion. }
\vspace{-0.0cm} 
\label{f-4}
\end{figure*}

We now illustrate that it is possible to further reorganize 
the TCA algorithm to achieve parallelization. For this we 
will use Message Passing Interface (MPI) parallelization.\\

\textit{a. PTCA scheme:} In Figure \ref{f-3}, a one-dimensional 
lattice example is used to illustrate the parallelization scheme 
of TCA. In the figure we show a $N=8$ site system with a $N_c=4$ site traveling cluster indicated by a rectangle. The update site is marked in red. Let us start with some initial values for the classical DOF 
at all the sites. The sites where an update has not yet been 
proposed are displayed in light blue. The rows from top to 
bottom indicate the different steps of a single MC system 
sweep where the update site traverses from left to right 
sequentially. The sites where the update have been attempted 
are colored in blue. Here for simplicity of presentation, we discuss the case where the `update site' is the site on the extreme left. Other choice of update sites are discussed in Sec VII. From the figure it is easy to see that assuming the MC system sweep starts from row one, the clusters in the first five rows do not depend on the update of the 
previous row. Rows six, seven, and eight depend on the outcome 
of the update attempts at sites one; one and two; and one two 
and three, respectively. In general, there are $N-N_c+1$ clusters 
of the former kind and $N_c-1$ of the later kind. We refer to the 
later kind as ``boundary clusters''.

In TCA the cluster diagonalization involved in all of the 
eight steps are carried out sequentially. The obvious way 
to parallelize the TCA is to diagonalize the independent 
sets of clusters in parallel. The simplest strategy for 
this is to divide the MC system sweep into two blocks, 
$L^{\prime}_1$ and $L^{\prime}_2$, each consisting of 
four of the eight steps of the MC sweep. It is easy to 
see that in this way all the four clusters in the block 
$L^{\prime}_1$ can be diagonalized on four processors 
simultaneously. Once the updated results for $L^{\prime}_1$ 
are received, they are used to generate the clusters for 
the $L^{\prime}_2$ block which can now be diagonalized in 
parallel. Thus the computation cost is $2N_c^3$ rather 
than $NN_c^3$ as in TCA. For a $d$ dimensional cubic system, 
the cost of PTCA scales as  $2^d N_c^3$. 
The $2^d$ factor comes from the correct accounting of 
all boundary clusters that can not be diagonalized 
simultaneously. This still is very advantageous as 
compared to the $NN_c^3$ scaling of TCA.

In  our approach, the one dimensional global system 
is broken into two blocks, each having $\frac{N}{2}$ 
number of sites. In two dimensions, the global square 
system is broken into four blocks and into eight in 
three dimensions.

\textit{b. PTCA flowchart:} We now discuss the implementation 
of PTCA. The flowchart is presented in Fig.~\ref{f-4} and it 
is discussed below. For simplicity we discuss specifically the one 
dimensional case, but a 
generalization to higher dimensions is straightforward.

For PTCA we will use $N_P+1$ processors with ranks 0 to $N_P$. 
As discussed below, of these the rank=0 processor is 
the master and is involved only in receiving and sending 
information, while the rest $N_P$ processors are the ones 
that will be used for diagonalization of clusters.
As in the TCA case, we define $S[~]$ as the array holding 
all the classical DOF for an $N$ site one dimensional system. 
As discussed above, we divide the system into two blocks, 
the loop label running over the blocks is ``K''. Within each block, 
another loop, labeled by ``I'', runs from one to $N_S/N_P$. 
Here $N_S$ is the number of sites within a block and we ensure 
that $N_S/N_P$ is an integer. Note that if $N_S=N_P$ 
the clusters built at all the $N_S$ sites can be diagonalized 
in one go. In PTCA the ``update site'', denoted by $P$, is a 
function of $N_S,~N_P$, and the rank of the processor on 
which the cluster built around $P$ is to be diagonalized. 
Thus the update site $P$ is denoted by  $P(K,I,Rank)$ in the 
flowchart. $C[P]$ is the cluster built around the update site 
$P(K,I,Rank)$. $H_{full}\{S[~]\}$ and $H_{clus}\{C[P]\}$ have 
definitions similar to that for TCA.

The MPI commands used are standard \cite{mpi} and will 
not be repeated here in detail. 
We use \texttt{MPI\_Init, MPI\_Comm\_size, MPI\_Comm\_rank} 
to allocate and assign labels (ranks) to $N_P + 1$ number 
of processors. The ranks of the processors range from $0$ to $N_P$.
\begin{enumerate}

\item Loop over blocks K (= $ 1,~2$) for our one dimensional example. 
\item Loop over I goes over $1,2 ,.. ,N_S/N_P$. 
\item For each I, assign the construction of the cluster 
around the ``update site'' $P(K,I,Rank=R)$ and the subsequent 
update procedure to the processor with $Rank=R,~(R>0$). The $N_P$ 
such assignments are depicted with small gray $N_P$ rectangles 
in Fig.~\ref{f-4}.

\item For a processor with $Rank=R,~(R>0)$, $C[P]$ will read 
the relevant $N_c$ site classical DOF data from $S[~]$. It will 
then diagonalize $H_{clus}\{C[P]\}$ before and after proposing 
an update for the classical variable at $P$. If accepted, 
the update of the $P^{th}$ site is sent to processor with 
Rank=0 using \texttt{MPI\_SEND}. This is shown in the expanded 
gray rectangle on the left of Fig.~\ref{f-4}.

\item Rank=0 processor receives update from all other processors 
with ranks 1 to $N_P$ using \texttt{MPI\_RECV} and suitably 
modifies the $S[~]$ on Rank=0 processor.

\item The ``I'' loop ends.

\item The Rank=0 processor broadcasts the modified $S[~]$ to all 
the processors using \texttt{MPI\_BCAST}, once updates from 
all processors have been received.

\item End the loop K.
\end{enumerate}

The parallelization scheme holds for any dimensions, as 
long as $N_c < N$ which we guarantee by definition. For 
the simplest case of $N_S=N_P$, the estimated total cost 
of $P_1$ MC sweeps with $P_2$ full system diagonalization 
for output calculations is $P_1 2^d N_c^3 + P_2 N^3$. 
This is a huge improvement in performance compared to 
TCA for which the computational cost for the same 
would be $P_1 N\times N_c^3 + P_2 N^3$. The improvement 
is significant when $P_1$ is a very large number, which 
is always the case. In the next section we present actual 
results when $N_S>N_P$ that establishes that even in this 
case the reduction of numerical cost is significant. For a 
fixed $P_1$ and $P_2$, beyond a certain system size, the 
full system diagonalization will dominate the total 
computational cost for the PTCA if these are ``done on the fly''. 
We suggest saving configurations and performing the full system 
diagonalization separately. A strategy for this process 
using Scalable LAPACK is suggested in section VI.

\section{Numerical benchmarks}
Let us now discuss benchmarks 
comparing TCA with PTCA. For this purpose we will use the following spin-fermion Hamiltonian:
\begin{eqnarray} 
 H_{\rm Hubb-MF}&=&
-t\sum_{\langle i,j \rangle,\sigma}c^{\dagger}_{i,\sigma} c^{\phantom{\dagger}}_{j,\sigma}\\\nonumber
&+&\frac{U}{2} \sum_i(\langle n_i\rangle n_i-{\bold m_i}.{\bold \sigma_i})\\\nonumber
&+&\frac{U}{4}\sum_i({\bold m_i}^2-\langle n_i \rangle^2)-\mu\sum_i n_i.
\label{e1}
\end{eqnarray}

This Hamiltonian is the SU(2) invariant Hartree-Fock mean field Hamiltonian for the Hubbard model. We have recently established  \cite{hubb-mcmf} that if the mean field expectation values in $H_{\rm Hubb-MF}$ are treated as classical variables and annealed via a classical MC process involving a slow reduction of the temperature, then the finite temperature results for all the observables we tested agree qualitatively and often quantitatively with Determinant Quantum Monte Carlo. 
In this model ${\bold m}_i$ is the mean field magnetization at the $i^{th}$ site 
and it is treated as a classical vector. $t$ is the hopping parameter 
and $U$ is the Hubbard onsite repulsion. We further set $\langle n_i\rangle=1$ 
for the case of half filling. This model that involves free electrons interacting 
with the classical spins defines our spin fermion model. We will present results 
in two and three dimensions and compare with those obtained using ED+MC 
and TCA in our earlier work~\cite{hubb-mcmf}. The methods used 
in [\onlinecite{hubb-mcmf}] have also been independently derived 
and applied in the context of the Hubbard model on an anisotropic triangular 
lattice \cite{triangular} and on geometrically frustrated face centered 
cubic lattices \cite{fcc}. Earlier similar approaches 
but for the attractive Hubbard interaction (negative $U$) where reported
in Refs.\cite{BdG0,BdG1,BdG2,BdG3,tarat1,tarat2,tarat3}. 

For the results presented in the present paper, the TCA cluster size used is $N_c=4^3$ in three dimensions and $N_c=4^2$ in 
two dimensions. We also checked the  independence of our results 
to variations in the traveling cluster size.
\begin{figure}
\centering{
\includegraphics[width=7cm, height=12cm, clip=true]{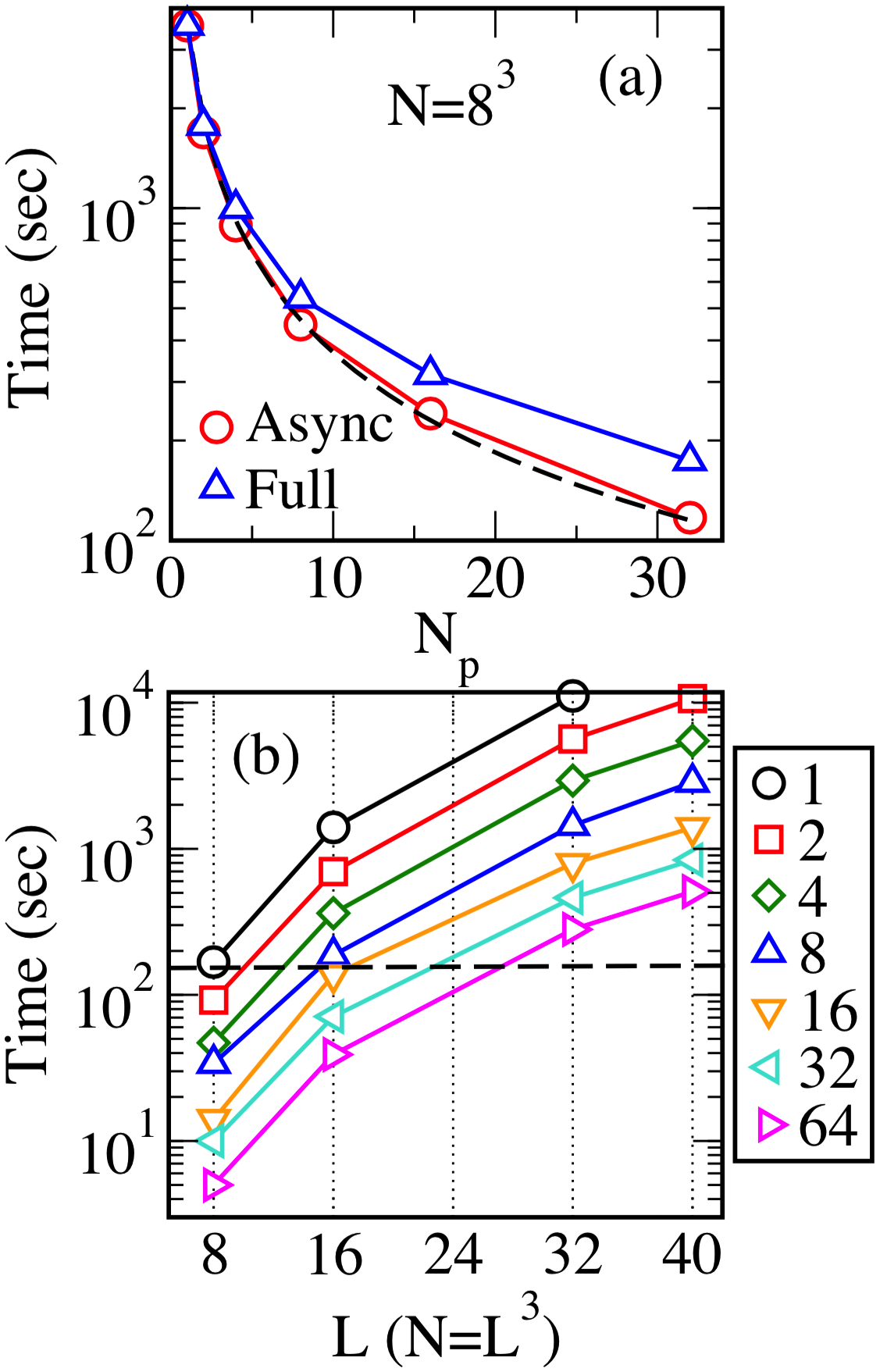}}
\caption{ (a) Time required for asynchronous ED and full ED with message passing 
against the number of processors $N_P$ at a fixed $U=8.0t$. See text for definitions. 
The data is presented for a $N=8^3$ lattice using 2000 MC system sweeps at a 
fixed temperature. The dashed line is the plot of $1/N_P$; (b) Time needed 
for 200 system sweeps for the full calculation with message passing 
plotted against $L$, for a system size $N=L^3$. The data for different 
number of processors, $N_P$, are shown. $N_P$ values are indicated on 
the right. The dashed line indicates that the computational cost for solving  
a $N=32^3$ system using PTCA with 64 processors is almost the same as 
the time needed using TCA on an $8^3$ system with a single processor. 
Calculations were done using multiple Intel Xeon E5-2670 processors 
which have eight cores with base frequency of 2.6 Ghz and 8 GB RAM per processor.}
\vspace{-0.0cm} 
\label{f-5}
\end{figure}

For a $L\times L\times L$ system in three dimensions, with a total number of sites 
$N=L^3$, the matrix size of the Hamiltonian for a given configuration of 
classical fields $\{{\bold m}\}$ is $2N \times 2N$. The factor of two comes 
from the two spin species of the fermions. Figure~\ref{f-5} is the main numerical 
result that establishes {\it (i)} the efficiency of PTCA over TCA and {\it (ii)} 
the dependence of the performance of the PTCA on the number of processors $N_P$. 
In (a) we display the bare time, without focusing on measuring physical results, 
for a $N=8^3$ lattice in three dimensions with cubic geometry. We further 
choose $N_c=4^3$. We have performed $2000$ MC system sweeps at a fixed temperature, 
that amounts to $8^3\times 2000$ or $1.024\times 10^6$  exact diagonalizations 
of $2\times 4^3$ matrices defining the traveling clusters. 
These are performed using the PTCA approach and employing  
different numbers of processors. The corresponding time needed 
is plotted in blue against $N_P$. Since large $N_P$ increases 
the communication time between the processors, for comparison in (a) we have also shown 
the time required to diagonalize the same number of matrices but with no interprocessor 
communication, labeled as asynchronous. This is indicated in red. Also the curve $1/N_P$ 
(the dashed line) establishes that the time needed for the asynchronous ED varies 
as $1/N_P$ within the PTCA scheme. When the processors communicate 
(using \texttt{MPI\_SEND}, \texttt{MPI\_RECV} and \texttt{MPI\_BCAST}), 
the time increases with $N_P$. But adds only a few seconds of additional 
time even for $N_P=32$. This is labeled as Full in Fig.~\ref{f-5}~(a).

In the previous section we had estimated that the cost of a system sweep 
in PTCA is  $P_1 2^d N_c^3$, for $P_1$ diagonalizations of the traveling 
cluster. However, this assumed that all independent traveling clusters in 
one block can be diagonalized simultaneously. Since the system sizes can 
be very large, this is seldom possible. As a result  only a fraction of 
traveling clusters in one block can be diagonalized simultaneously. It is 
easy to check that this would lead to the $1/N_P$ dependence seen in Fig.~\ref{f-5}~(a) 
apart from the additional processor communication time. In (b) we show the time needed 
for 200 MC system sweeps within PTCA against $L$, for a $N=L^3$ system. This is displayed 
for different $N_P$ values indicated on the right. From (b) it is clear that the time for 
the 200 system sweeps for a $N=8^3$ system with single processor (or within TCA) is almost 
equal to the PTCA cost for a $32^3$ system with $N_P=32$.

With this clear advantage of PTCA, in the next section we discuss some particular 
physics results and compare them with existing literature.

\section {Results for the Hubbard model in two \& three dimensions}

\begin{figure}[t]
\centering{
\includegraphics[width=8cm, height=8cm, clip=true]{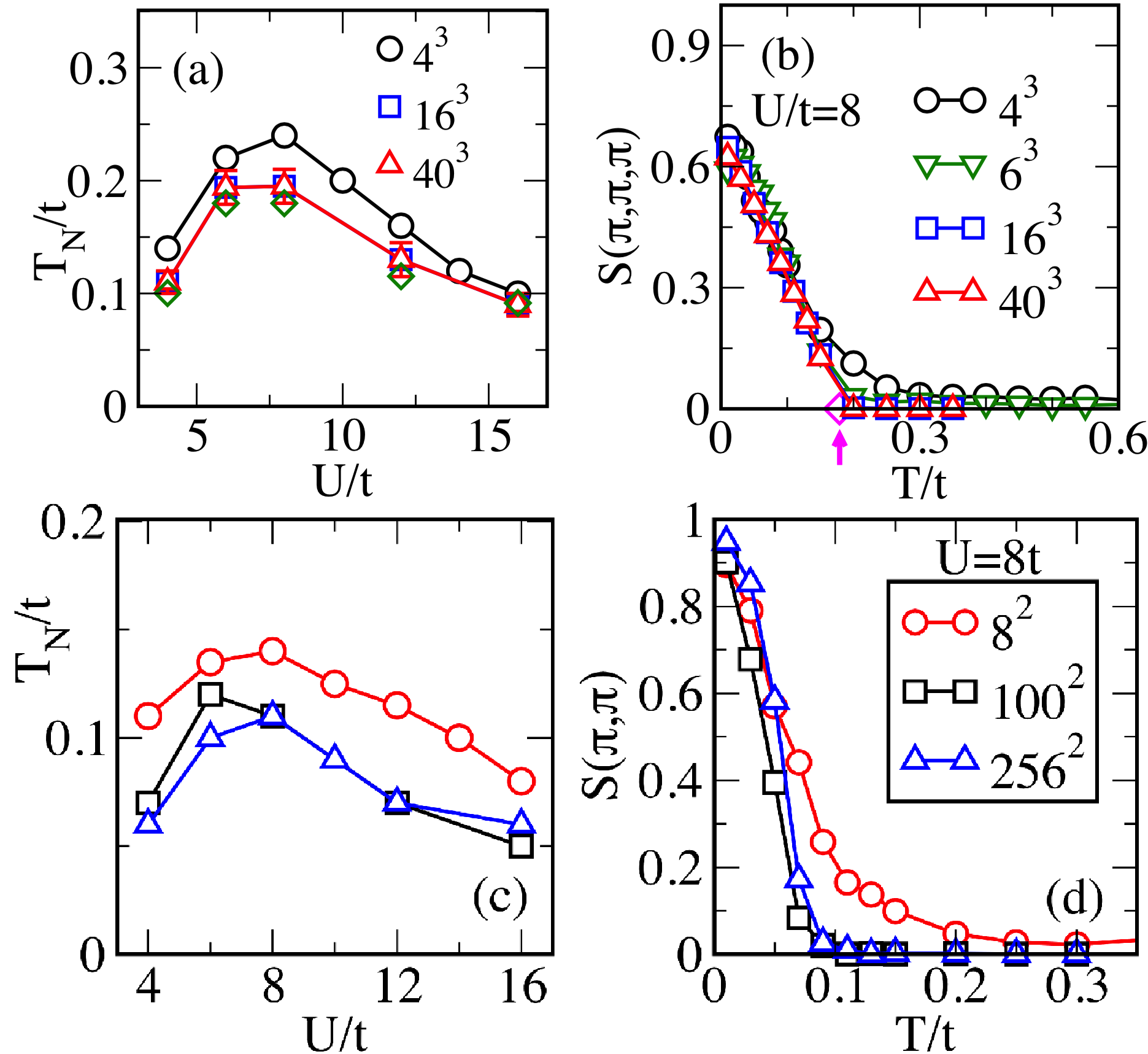}}
\caption{(a) $T_N$ vs. $U/t$ for system sizes with $4^3, 16^3$, and $40^3$ sites. (b) The spin structure factor S($\pi,\pi,\pi$) vs. temperature, for system sizes $4^3, 6^3, 16^3$, and $40^3$ at $U = 8.0 t$. The results are obtained using a $4^3$ traveling cluster and 16 processors. The data is averaged over 200 configurations obtained from 2000 MC steps for the cases  $N=4^3$ and $N=6^3$, while only 1250 MC steps were used for $N=16^3$ and $N=40^3$ with an average over 50 configurations. The magenta arrow indicates the thermodynamic $T_N$ 
in this case. (c) and (d) present the corresponding results in two dimensions. (c) $T_N$ vs. $U/t$ and (d) the corresponding S($\pi,\pi$)'s for system sizes $8^2, 100^2$, and $256^2$. The results are obtained using a $4^2$ traveling cluster and 16 processors.}
\vspace{-0.0cm} 
\label{f-6}
\end{figure}

In Fig.~\ref{f-6} we discuss the magnetic properties of the Hubbard model in two and three dimensions by studying Hamiltonian Eq.~(1) at finite temperature using PTCA. In our recent work we have extensively studied this system using EDMC and TCA \cite{hubb-mcmf}.
At half filling the N\'eel temperature, $T_N$, has a non monotonic dependence on $U$. These results are presented here in two and three dimensions. In Fig.~\ref{f-6}~(a) we plot $T_N$ against $U$ for three different system sizes, $4^3$, $16^3$, and $40^3$. These are all obtained using PTCA. The $4^3$ results are identical to the $4^3$ results in our earlier work. 
For the larger system sizes studied here we find that $T_N$ converges and it has a weak dependence on the finite size of the system. We emphasize that until now in the literature there have been no results for spin-fermion models employing such large number of sites. We use these large system values of $T_N$ to perform finite size scaling. 
In Ref.~\onlinecite{hubb-mcmf} we had established that the magnetic structure factor obtained with the TCA agrees with the ED+MC data at all temperatures. This indicates that finite size effects associated with the cluster size do not affect the finite temperature evolution of the magnetic state appreciably. Thus, the finite size scaling using TCA or PTCA is justified. 

For these results on finite systems, the bulk $T_N$ estimates are obtained  
by an inspection of the $S(\pi,\pi,\pi)$ data shown in (b). Information 
regarding the N\'eel AFM order is obtained from the magnetic structure 
factor for the ${\bold m}$ variables,
\begin{equation}
S({\bold q}) = \frac{1}{N^2} \displaystyle\sum\limits_{i,j} e^{i {\bold q} \cdot ({\bold r}_{i}-{\bold r}_{j} )} 
\langle {{\bold m_i} \cdot {\bold m_j}}\rangle, 
\end{equation} 
where ${\bold q}=\{\pi,\pi,\pi\}$ is the wavevector of interest.

Then, assuming that the correlation length  $\xi(T_N(L)-T_N^{Thermo})=aL$ on a $L^3$ system, and given that $\xi(x)\propto|x|^{-\nu}$, one arrives at the scaling form, 
$T_N(L)=T_N^{Thermo}+bL^{1/\nu}$. Here, $L$ denotes data from a system 
size $N=L^3$. We plot the finite system N\'eel temperatures against $1/L$ 
and use $T_N^{Thermo}$, $b$, and $\nu$ as fitting parameters. Details of this process 
are presented in our earlier work, and here in Fig.~\ref{f-6}~(a) we simply present the 
results (green diamonds). We have found that indeed both the $16^3$ and $40^3$ results 
converge to the true thermodynamic N\'eel temperature. For the antiferromagnetic 
structure factors for different system sizes in (b), we find that the PTCA 
results for $N>16^3$ are virtually identical and appreciably better than the $4^3$ and $6^3$ results which have non-negligible finite size effects. The arrow in (b) indicates the thermodynamic $T_N$ as obtained 
from finite size scaling for the case $U/t=8$.

In (c) and (d) we show the corresponding results in two dimensions. 
Here, as it is well known, in principle the Mermin-Wagner theorem 
establishes that there is no true $T_N$ in two dimensions for an $O(3)$ magnet. 
However, this theorem is valid only for short-range 
spin-spin interactions. In 
our case, the integration of the fermions leads to effective spin-spin interactions 
at all distances, although the rate of the decay of the couplings with distance is 
unknown. Panels (c) and (d) indicate 
that the $N=100^2$ and $N=256^2$ results, while 
significantly lower than the $8^2$ result, are very 
close to each other suggesting convergence. However, this subtle matter 
requires further discussion and larger clusters to be fully
understood and our goal in this section is merely to check the performance of the
proposed PTCA method. The clarification of the validity of the Mermin-Wagner
theorem for spin-fermion models is left for the future.

%Finite size analysis of the two dimensional data, not performed here, will be of interest, however.

\section{Diagonalization of full system}
In this section we will discuss the strategy for diagonalizing large full systems 
to calculate fermionic observables that in principle require all eigenvalues and all eigenvectors
for each configuration of classical variables. 
In the PTCA scheme, given the large matrix sizes for the full system, 
we find that it is best to first simply anneal the classical variables, 
then store many equilibrium configurations at each temperature generated during the
Monte Carlo process, and then at the end perform full system diagonalizations to 
calculate the fermionic observables separately. In the special cases 
where we are interested only in the correlation among classical variables of course
we can certainly measure those correlations for each MC configuration.
But for the fermionic observable 
cases that require, e.g., full Green functions 
we suggest using Scalable LAPACK for the parallel diagonalization of the full system
using the equilibrium configurations.

\begin{figure}[t]
\centering{
\includegraphics[width=8cm, height=5cm, clip=true]{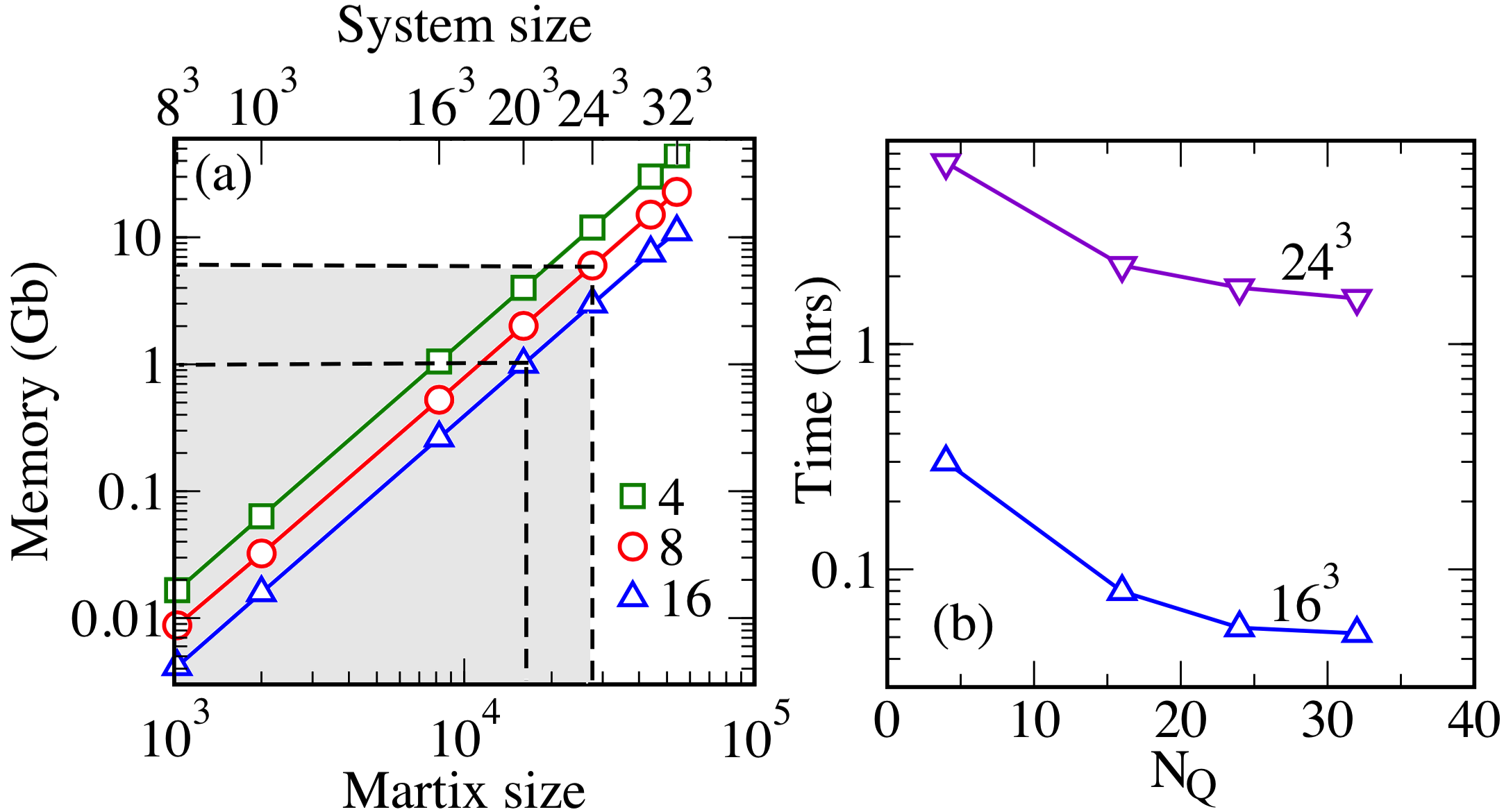}} 
\caption{Memory and time needed for a single diagonalization using Scalable LAPACK. In (a) we show the RAM (in Gb) per processor that is required to store the arrays necessary to diagonalize a double complex Hermitian matrix. The gray region is defined by a limit of 4Gb RAM per processor, which can diagonalize a Hamiltonian for a $24^3$ system (27648 dimensional matrix) with 8 processors. The data is also presented for 4 and 16 processors; (b) the time needed for a single diagonalization for $16^3$ and $24^3$ systems vs. the number of processors $N_Q$ used in Scalable LAPACK.}
\vspace{-0.0cm} 
\label{f-7}
\end{figure}

Let us assume $N_Q$ are the number of processors used for diagonalizing the large 
matrices employing Scalable LAPACK. Figure~\ref{f-7}~(a) shows the memory required 
to store all of the arrays that are necessary to diagonalize a large double 
complex hermitian matrix. It should be noted here that the Hamiltonian as one 
complete array is never created on an individual processor. Instead, the Hamiltonian 
is evenly spread out in blocks among all of the processors. This greatly reduces the 
total RAM required as well as the RAM per processor \cite{slug}. The total RAM and 
number of processors for a given system is a constant, therefore the ratio of RAM 
per processor is a fixed quantity. For example, in the traveling cluster used for 
these calculations, every job (run) submitted is allocated 2 Gb per processor. If 
one job uses more RAM than this, some processors can not be used since they do not 
have memory available to them. Therefore, when diagonalizing large matrices 
using the number of processors that approaches this fixed ratio will optimize 
the CPU time and memory usage. In (a) we see that, as expected, 
the memory requirement 
grows with matrix and system size, but reduces with increasing $N_Q$. 
The $N_Q$ values 
of 4, 8, and 16 are indicated in the figure. The gray region in (a) is where memory 
needed per processor is 4Gb or less. For typical computational resources of multicore 
workstations this is easily available. This requirement corresponds to a system size of 
about $24^3$ sites.

The other issue is the time needed for diagonalization. In Fig.~\ref{f-7}~(b) we present the typical time needed for single diagonalization corresponding to $N=16^3$ and $24^3$ system sizes or matrix sizes $8192 \times 8192$ and $27648 \times 27648$, respectively. The results are for $N_Q=4$, $16$, $24$, and $32$ processors. 
In both cases the time gain is quite significant with increasing $N_Q$. 
If all the configurations over which the output quantities are to be averaged 
at a fixed temperature are calculated in parallel, then for $N_Q=8$ 
we require only about 0.1 hours of additional computation time for a $N=16^3$. 
For $N=24^3$ the additional time is about 2 hours. The additional time goes down 
further for larger $N_Q$. High end workstations and small clusters should easily 
be able to supply the resources needed for such system sizes.

\section {Discussion}

\begin{figure}[t]
\centering{
\includegraphics[width=8cm, height=4.5cm, clip=true]{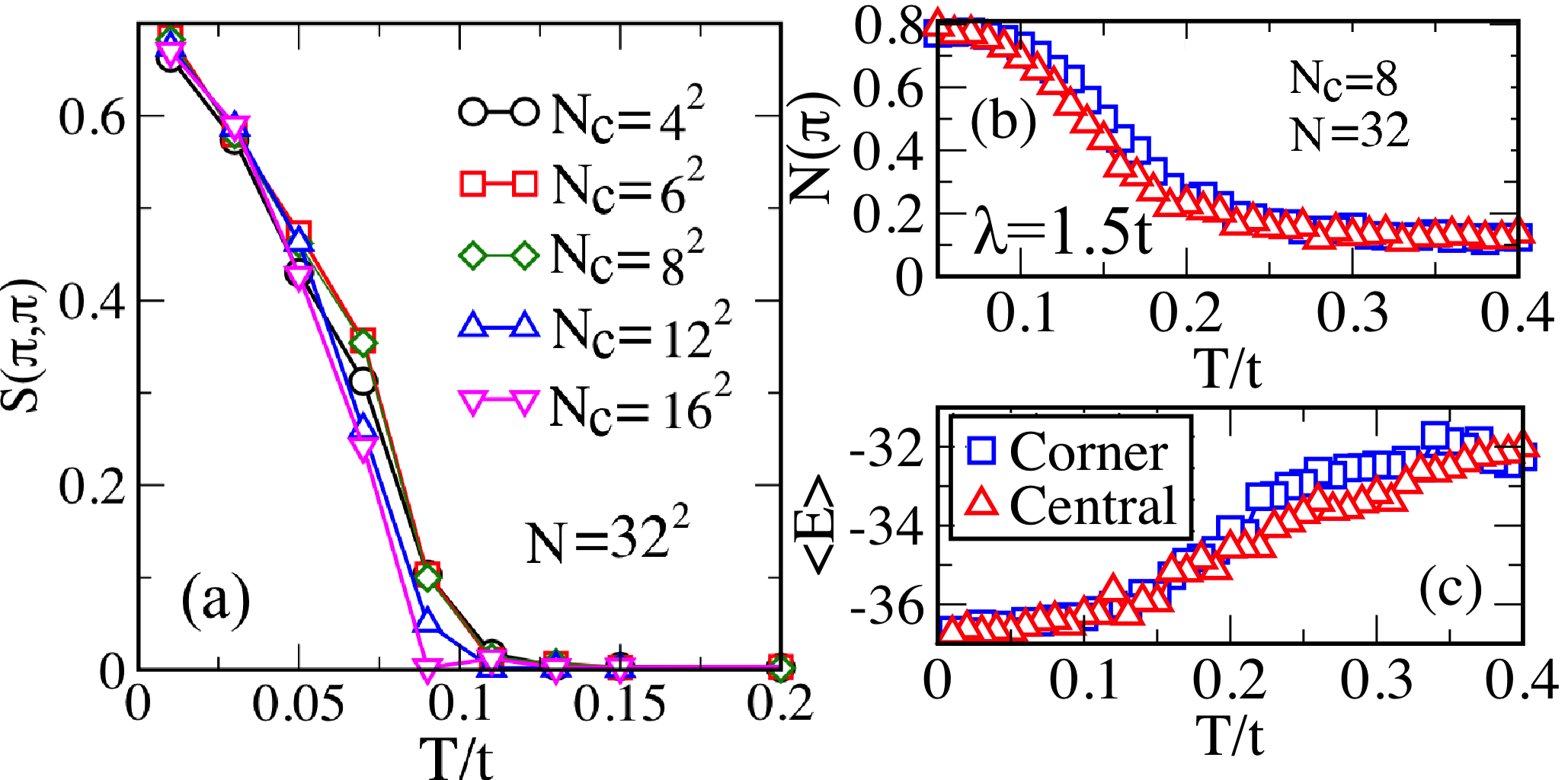}}
\caption{(a) Antiferromagnetic order, S($\pi,\pi$), for the two dimensional Hubbard model at $U/t=8$ using a $N=32^2$ lattice with difference traveling cluster sizes as indicated. (b) The one dimensional checkerboard charge order parameter vs. temperature for the adiabatic Holstein model. (c) shows the corresponding average energy of the system with temperature. The data is presented for two schemes indicated as `central' and `corner', see text for discussion. The parameters for (b) and (c) are the same.}
\vspace{-0.0cm} 
\label{f-8}
\end{figure}

In this section, we will discuss two related numerical issues and provide an estimate for the cost of solving spin-fermion models derived from multiorbital Hubbard models:

\textit{a. Cluster size effects:} The first is the dependence of results on the traveling cluster size. In Fig.~\ref{f-8}~(a) we show the antiferromagnetic structure factor vs. temperature for a lattice with $32^2$ sites employing different sizes for the traveling clusters, using the same mean field Hubbard model discussed before. 

While small cluster sizes are good enough to capture the long range order as well as the rough location of the transition temperature for this model, the finite cluster sizes introduce finite size effects of its own. To reduce these, one needs to employ larger traveling clusters. In Fig.~\ref{f-8}~(a) we see that finite size effects in $T_N$ reduce rapidly with larger clusters, $N_c=12^2$ and $16^2$, for the same fixed system size. Furthermore in physical  problems where there is long wavelength order, large $N_c$ would be crucial to capture the correct phases. 
In TCA, the linear depencence of the numerical cost on the system size limits $N_c$ to $8^2$. Larger $N_c=12^2, 16^2$, results are only possible within the current scheme. We would like to emphasize that this is an additional significant improvement over TCA.

\textit{b. Choice of the update site:} In Fig.~\ref{f-2} we had displayed a scheme for setting up the PTCA. There we had chosen the leftmost site of the cluster as the site where the update is attempted. Choosing this leftmost ``update site'' was mainly for convinience. Here we briefly demonstrate the effect of choosing other update sites. The parallelization, of course, applies to any such choice.
%As discussed in section 2, in TCA any site in the traveling cluster is a valid choice for the update site, as long as that same site is chosen for all clusters. 
%This later requirement ensures that updates are attempted at all sites of the lattice. 
In Figs.~\ref{f-8}~(b) and (c) we show the comparison of results for different choices of the update site.
For this purpose, we study the adiabatic Holstein model in one dimension at half filling. The Hamiltonian for  this model is 

\begin{eqnarray}
H_{\rm Hol}&=&-t\sum_{\langle i,j \rangle,\sigma}c^{\dagger}_{i,\sigma} c^{\phantom{\dagger}}_{j,\sigma} \\\nonumber
&+&\sum_{i}(\lambda x_i-\mu)(n_{i}-1 )+K/2\sum_{i}x_i^2,
\label{e2}
\end{eqnarray}
where $n_i=(n_{i,\uparrow}+n_{i,\downarrow})$. 
In the particle-hole symmetric adiabatic Holstein model, the classical variables $\{x\}$ at every site denote the static lattice distortions. $\lambda$ is the electron lattice coupling and $K$ regulates the elastic cost of the lattice deformation. 
In this model the goal is to determine the optimal configuration of the  $\{x\}$ variables that minimizes the free energy. At half filling the model exhibits a checkerboard charge order together with large and small lattice distortions \cite{holstein}. The charge order can be probed by plotting the structure factor for the classical variables. This is defined by
\begin{equation}
N(q) = \frac{1}{N^2} \displaystyle\sum\limits_{i,j} e^{i {q} ({r}_{i}-{r}_{j} )} 
\langle {{ x_i} \cdot { x_j}}\rangle, 
\end{equation} 
where $q={\pi}$ is the wavevector of interest.

In our study two schemes were used: scheme `1' where the update site is 
the leftmost site of the traveling cluster, and scheme `2' where the $(N_c/2+1)^{th}$ site is the update site. In the one dimensional study with $N=32$ and $N_c=8$, the sites 1 and 4 are the choices for the update site and the two schemes are refered to as `corner' and `central', respectively. We do not present the details of the algorithm for `central' scheme here, which is very similar to the earlier scheme. We just mention here that one needs to choose a different way of distributing which clusters are to be diagonalized in parallel. The numerical advantage is comparable in both schemes.

In Fig.~\ref{f-8}~(b) we study the correlation between the classical variables. $N(q=\pi)$, as defined before, is plotted as a function of temperature. At low temperature an alternating large-small pattern generates a peak at $q=\pi$ in the charge structure factor. As seen in the figure, the results from both schemes match with each other. In addition in (c) we show the average energy with temperature, which also agrees over a wide temperature range.

\textit{c. Numerical cost for multiorbital Hubbard model:}
To derive the general formula for numerical cost of PTCA for a multiorbital Hubbard system, we first note from Sec. IV, that we had divided the system into $2^d$ blocks. Thus $N_S=N/2^d$, where $N_S$ is the number of sites in a block. Secondly, since we build a cluster around each of those $N_S$ sites in a block, the time taken for a MC system sweep is simply the cost of a single cluster diagonalization times the number of blocks. To do so, however, requires us to diagonalize all $N_S$ clusters in a block simultaneously. This would require $N_S$ processors. Typically, for large systems, the number of processors $N_P$, is much smaller than $N_S$. In such cases only $N_P$ number of clusters in a block can be diagonalized simultaneously. Thus the cost to complete the diagonalizations of all $N_S$ clusters in a block would be $N_S/N_P$ times the cost of diagonalization of a single cluster.

From these, it is easy to deduce that the cost for $P_1$ MC steps in the PTCA as discussed in section IV, $P_1 2^d N_c^3$, can be written as $P_1 (N_c)^3\times\frac{N}{N_P}$. As a consequence, the cost for $P_1$ MC steps for $N_O$ orbitals (with two spins per orbital), would be $P_1 (2N_O N_c)^3\times\frac{N}{N_P}$.
From this expression it can be shown that if $N=N_P$ then the cost of PTCA is the cost of $P_1$ cluster diagonalizations. If $N_P=1$, then the cost grows linearly with $N$, which is precisely the case for TCA. Finally for a general $N_P$, the cost scales as $1/N_P$.

\section{Conclusions}
In conclusion, we have provided a reorganization of the TCA algorithm that allows for a straightforward parallelization. To test the method, we have presented results for the Hubbard model in two and three dimensions treated in the mean field approximation and for the Holstein model with classical lattice distortions in one dimension. 
A comparison with earlier work clearly shows that the PTCA approach can produce reliable results on very large lattices. Apart from accessing large system sizes for the case of the single orbital Hubbard model, the new approach will facilitate the study of finite temperature effects in multiorbital Hubbard models, treated in the mean field approximation, where the large orbital degeneracy (up to five orbitals in models for iron superconductors) severely limits the number of sites that can be solved employing ED+MC, even when including the TCA improvement.\\

\section{Acknowledgments}
We acknowledge the use of the Newton cluster at the University of Tennessee, Knoxville, where all the numerical work was performed. A. M., N.P., and C.B. wrote the computer codes, and gathered and analyzed the results. They were partially supported by the National Science Foundation under Grant No. DMR-1404375. E.D. guided this effort and contributed to the writing of the manuscript. E.D. was supported
by the U.S. Department of Energy, Office of Science, Basic Energy Sciences, Materials Science and Engineering Division.
\bibliography{pca2.bib}
\end{document}